\documentclass{kluwer}
\usepackage{graphicx}
\newcommand\fm{\mbox{$.\!\!^{\mathrm m}$}}%

\begin{document}
\begin{article}
\begin{opening}
\title{\bf OPTICAL PHOTOMETRY OF V4334 Sgr\\ 
(SAKURAI'S OBJECT)}

\author{HILMAR W. \surname{DUERBECK}\email{hduerbec@vub.ac.be}}
\institute{WE/OBSS, Vrije Universiteit Brussel, Pleinlaan 2, B-1050 Brussel,
  Belgium}

\runningtitle{OPTICAL PHOTOMETRY OF V4334 Sgr}
\runningauthor{HILMAR W. DUERBECK}

\begin{ao}
Hilmar W. Duerbeck\\
WE/OBSS\\
Vrije Universiteit Brussel\\
Pleinlaan 2\\
B-1050 Brussel\\
Belgium
\end{ao} 

\begin{abstract} 
The optical photometric evolution of the final helium flash object V4334 Sgr 
from 1994 to 2000 is described. The rise to optical maximum (1994 -- 1996) 
is characterized by a continuous increase of color indices, indicating a 
slowly expanding, cooling pseudo-photosphere. This photosphere became 
stationary in 1997. In the following years, the object underwent brightness
declines of increasing strength, which are similar in character to the ``red
declines'' of R CrB stars. The fading of V4334 Sgr is more dramatic than any
brightness decline of an R CrB star: at present, only 1/100000 of the visual 
light reaches the observer. Most radiation is absorbed by the dust envelope 
and re-radiated in the infrared. The sparse optical data of 2000 show that 
the obscuration has not increased in strength any more. The light curve of 
V4334 Sgr is similar to that of the final helium flash object V605 Aql which 
erupted in 1919. 

\end{abstract}


\end{opening}

\section{Introduction}

From the observer's standpoint, 
objects undergoing a final helium flash are rare events. 
Each year, a handful of classical novae are discovered, 
whose eruptions are triggered when the hydrogen-rich 
material accreted onto a C-O or Ne-Mg white dwarf from a 
close companion star reaches a critical mass. Such an 
event may repeat with a timescale of several thousand years.
In contrast, a final helium-flash candidate accretes helium from the 
outer hydrogen-burning shell, and builds up a critical helium mass
above its degenerate C-O core. It can happen that the last 
helium flash occurs when the star is still in its giant phase. 
Thus the event remains unnoticed by an outside observer, 
and no critical amount of He is accreted during the subsequent 
evolution of the object along the track of central stars of planetary
nebulae towards the white dwarf region. 

Iben, Tutukov and Yungelson (1996) have estimated that approximately 15\% 
of intermediate mass stars experience an {\it observable} 
final helium flash. In spite of the fact that nova eruptions occur on
very special close binary stars, the repetivity of the nova event overwhelms
the singular event of a final helium flash that occurs on many ordinary 
stars on their way from the AGB to the white dwarf stage.
\begin{figure}
\begin{center}
\includegraphics[width=100mm,angle=0]{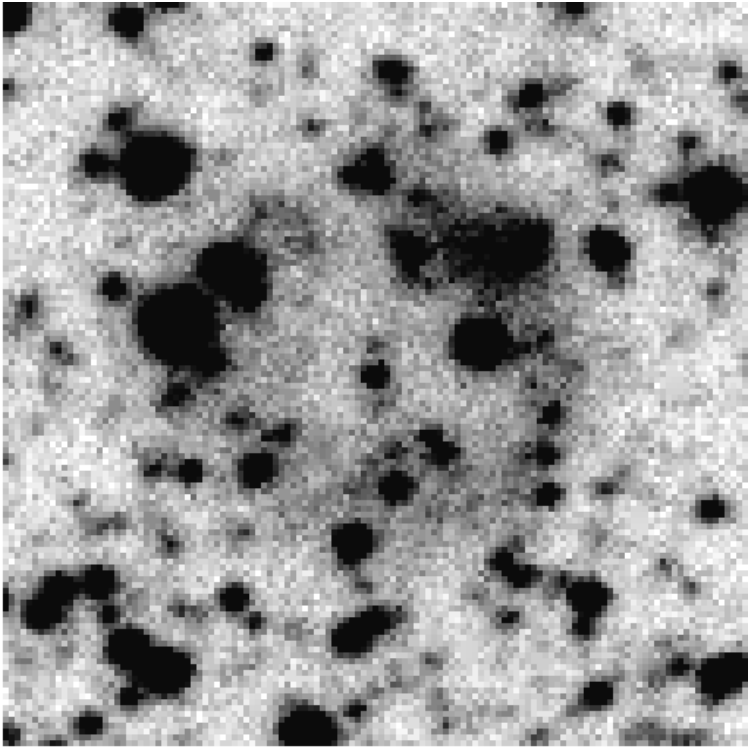}
\vspace*{-5cm}
\caption[]{
A direct image of V4334 Sgr, taken through a narrow-band H$\alpha$ filter 
with the 0.91\,m Dutch telescope at ESO  La Silla on 1999 March 26 by 
C.~Sterken. The exposure time was 1000 sec. The size of the field is 
about $1'\times 1'$. North is to the top, east to the left. The planetary 
nebula is clearly seen; the central object, at about $R = 17.7$, is 
flanked by a northern and a southern component with $R\sim 20^{\rm m}$.}
\end{center}
\end{figure}

A catalog of 277 galactic novae and related stars contains one object 
that was identified as a final helium flash (V605 Aql, alias Nova Aql 1919); 
the nature of a second object (CK Vul = Nova Vul 1670) remains 
unclear, and maybe a few more final helium-flash objects 
may hide among the faint, poorly observed novae. In any case, 
from the standpoint of the observer, final
helium flashes are about 100 times as rare as thermonuclear hydrogen
runaways.

It came as a pleasant surprise when on February 20, 1996, Yukio Sakurai 
discovered ``Sakurai's object'', which was first announced as a 
novalike object. It soon turned out to be such a rare final helium flash 
(Fig.~1). The discovery of Sakurai's object = V4334 Sgr permited
for the first time to follow the evolution of a final helium flash 
object with modern equipment, especially in the optical and IR range. 
The belated discovery (the object already had been bright during all 
of 1995) precluded early color and spectroscopic observations. A UV 
spectrum taken soon after discovery during the last weeks
of IUE (Gonzalez-Riestra, private communication) 
is unfortunately too noisy to reveal any unusual features.

\section{The pre-helium-flash object and the early rise}

Since the field of V4334 Sgr has a photometric magnitude scale down to
$21^{\rm m}$, interesting conclusions can be drawn from Schmidt plates 
taken for various sky survey. On 1976 May 30, the object appears at
$21\fm 0$ on a UKSTU $J$-plate, and it is fainter than 
$21\fm 5$ on the 1984 July 26 ESO-Schmidt $R$-Plate. Three subsequent 
$R$ plates taken with the UKSTU 
in July 1986, September 1989 and July 1991 show a feature which may be 
interpreted as a star (or a blend of stars) near $R=20\fm 7$. Thus 
we can exclude any strong photometric variability at minimum, and can 
limit the onset of the helium flash to a time after mid-1991.

\begin{figure}
\begin{center}
\includegraphics[width=100mm,angle=270]{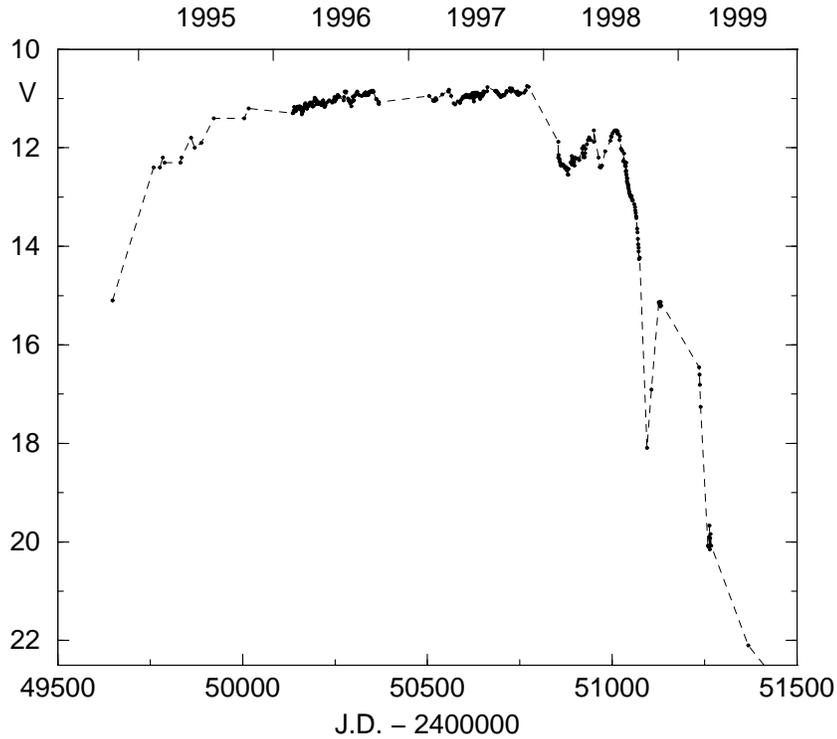}
\caption[]{
The $V$ light curve of V4334 Sgr, 1995-1999. The data of 1994 -- 1995 are the
photographic prediscovery observations of Takemizawa, converted to visual
magnitudes. Other observations were obtained with the 0.91\,m Dutch 
and the 0.2\,m Re$\rm \tilde{n}$aca telescopes. A few observations 
by Jacoby \& De Marco are also included.}
\end{center}
\end{figure}

Four film exposures taken 
by Takemizawa (1997) show V4334 Sgr at or below $m_{\rm pg} 
= 15.5$ in April -- September 1994, another one of October 22, 1994 below 
$m_{\rm pg} = 15.0$. The first exposure of 1995, on February 10, shows 
it at $m_{\rm pg}= 12.4$. The object was slowly rising in brightness during 
the subsequent weeks and months: by October 25, it had reached $m_{\rm pg} = 
15.5$, but it was only discovered after its conjunction with the Sun, on 
February 20, 1996 (Fig.~2).

\section{V4334 Sgr at optical maximum -- standstill,
  fluctuations and reddening} 

The actual discovery of V4334 Sgr in early 1996 was followed by a close 
monitoring
during the following years. We focus here on the {\it UBVRi} observations
carried out with the 0.91\,m Dutch telescope (1996 -- 1999)
and the 1.54\,m Danish telescope at ESO La Silla (1999 -- 2000), and
the $V$ observations with Liller's 0.2\,m f/1.5 Schmidt telescope at
Re$\rm \tilde{n}$aca (1996 -- 1998). Other series of optical photometry were
carried out by Guinan et al. (1998) and Arkhipova et al. (1999).

The 1995 -- 1996 visual light curve can be interpreted as produced by an 
object which radiates at almost constant bolometric luminosity.
This model assumes an instantaneous `switching on' of the nuclear 
energy source. The energy is partly used to lift the degeneracy and 
make the outer layers expand, and is partly radiated away at the
pseudo-photosphere whose position is determined by the actual mass loss
rate (``thick wind model'', see, e.g. Bath and Harkness 1989). The rise 
of the visual light curve may just be the shift of the radiation 
maximum of the source from the ultraviolet to the optical window
(a similar phenomenon is observed in the pre-maximum light curve of
classical novae, where the photospheric expansion is, however, 
more rapid by a factor $\sim 1000$). 

In 1996, the light curve was still dominated by a cooling of the photosphere, 
yielding a continuing fading at short and a brightening at long wavelengths 
(Fig.~3). Assuming a distance of 8 kpc, Duerbeck et al. (1997) derived an 
expansion of the pseudo-photosphere of $\sim 1$ km/s. This expansion 
slightly accelerates with time. Its extrapolation back to ``zero radius'' 
leads to a time of outburst around 1994 September 24.
 
\begin{figure}
\begin{center}
\includegraphics[width=100mm,angle=270]{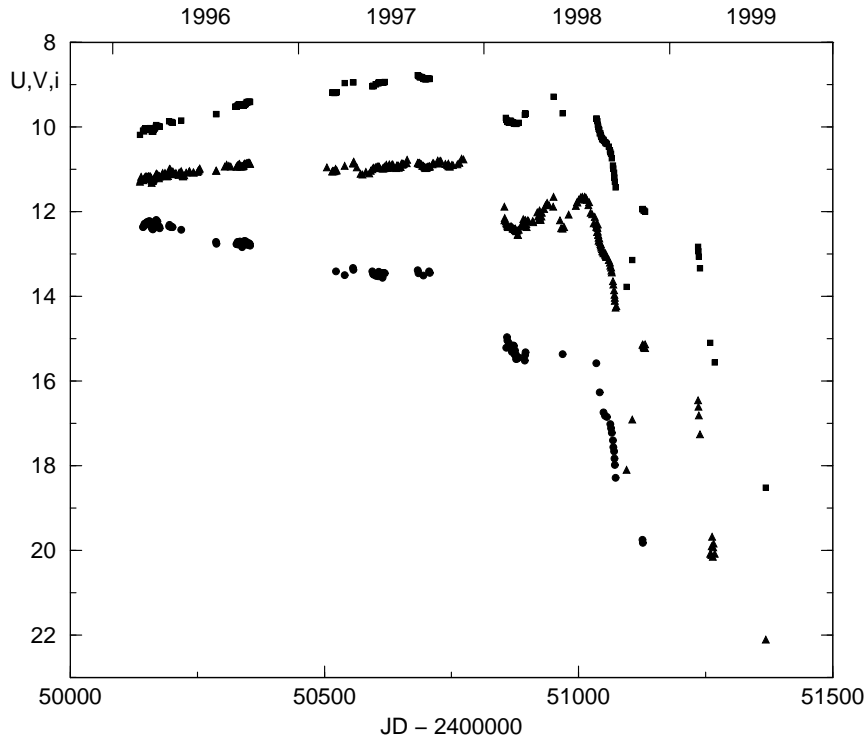}
\caption[]{
Multicolor light curve of V4334 Sgr. Only the $U$ (circles), $V$ (triangles) 
and $i$ (squares) light curves are
shown for clarity. The cooling in 1996, the standstill in 1997, and the onset
of dust formation in 1998 are clearly seen.}
\end{center}
\end{figure}

But even if the above model is not used, and it is assumed that 
the visual light curve of V4334 Sgr mimicks the total luminosity
evolution, i.e. a ``slow'' turn-on of the energy generation, the 
rise from minimum to maximum light must have taken only 2 to at 
most 5 years (the last value being extremely unlikely), according to 
the available pre-outburst record. This is much faster than in any 
of the calculated models (Iben et al. 1983, Iben and MacDonald 1995).

In 1997, the expansion had nearly come to a halt, yielding almost
constant color indices. Superimposed on the general brightness evolution were 
quasiperiodic fluctuations of increasing cycle length and amplitude. It seems
that several pulsation modes with periods between 63 and 8 days were excited. 
Models that show this type of pulsation werre developed by A.~Gautschy
and are described by Duerbeck et al. (1997). 
The pulsations have larger amplitudes at shorter wavelengths, e.g. for
J.D.mod. 50\,161 -- 169, the fluctuation at $(U-B)/(B-V) \sim 1$
(See Duerbeck et al. 1997). Note that ``red declines'' of R CrB 
and V854 Cen have $(U-B)/(B-V)$ slopes of 1.0 and 0.6, thus supporting 
Kimeswenger's (2001) conjecture that these fluctuations were already 
caused by miniature dust events. Compared to R CrB declines, the time scales 
of these mini-declines are very short, and their gradients are small ($\Delta
V/\Delta t \sim 0.02 $ mag/day.) Thus it appears more likely that V4334 Sgr at
maximum exhibits pulsations which are similar to those found in R CrB stars 
at maximum light. 

We assume that at this ``maximum'' phase of 1997, V4334 Sgr was not yet 
surrounded by noticeable amounts of circumstellar dust. The IR excess 
observed in 1996 -- 1997 corresponds to blackbody temperatures of 
3500$\rightarrow$2000~K, and may be caused (1) at earlier times by 
free-free radiation, and (2) at later times by carbon nucleation in shocks
that occur in the extended shell whick lie soutside the pseudo-photosphere. 
The observed Balmer decrement of the surrounding planetary nebula (Kerber et
al. 2000), the comparison of colors of V4334 Sgr with synthetic colors of
hydrogen-deficient stellar atmospheres, and the interstellar reddening of
field stars in the vicinity of V4334 Sgr leads to the assumption that
the the interstellar reddening is $E_{B-V} = 0.8$ 
(see also Duerbeck et al. 2000).

\section{The onset of dust formation}

1998 was dominated by dust events of increasing strength. When first observed 
in early 1998 after its conjunction with the Sun, the object was already one 
magnitude fainter than 80 days before. A weak recovery was followed by 
another weak depression, and afterwards a third very strong depression set
in. After that, observations became sparse (since the object had sunk below
the limiting magnitude of Liller's telescope); a few weak recoveries from
fainter stages are indicated, but the general trend was a decrease to 
$V \sim 22^{\rm m}$ in mid-1999. The photometric record for 2000 is sparse, 
and hampered by the fact that (a) the object is very faint in $V$, and (b)
the magnitude scales at these levels are not yet well calibrated. Observations
in February, April and July 2000 show the object fluctuating, but not any more
declining at a magnitude of $i=20.8 \pm 0.25$. The color indices were 
$V-R=0.5$ and $V-i=3.4$; the blueing of the color indices, already noted in
July 1999, seems to have continued, especially in $V-R$. A thorough discussion
will be postponed until more accurate photometry is available.

\section{The massive dust shell phase and the quest for the luminosity}

The evolution of V4334 Sgr in various color-color and brightness-color diagrams
(Fig.~4) is similar to the ``red declines'' that occur in R CrB variables, 
i.e. a general reddening of color indices described by a star which is 
completely obscured by a dust cloud.
Duerbeck et al. (2000) have compared the fading of V4334 Sgr with that of 
the R CrB variable V854 Cen, and find that they are quite similar. There are,
however, two major differences: While the deepest decline observed in an R CrB
variable was $8^{\rm m}$, V4334 Sgr has experienced a decline (in $V$) of
13 magnitudes, and appears to have stabilized (in 2000) at a very faint 
magnitude, with no indication for a quick recovery.
\begin{figure}
\begin{center}
\includegraphics[width=110mm,angle=270]{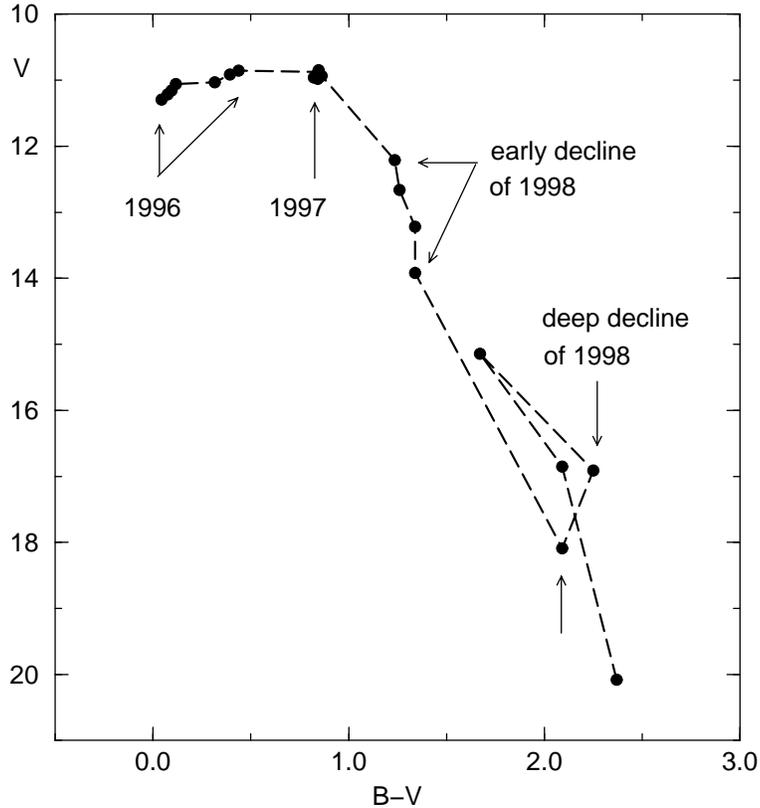}
\caption[]{The $V$ vs. $(B-V)$ diagram of V4334 Sgr. The $(B-V)$ values are
  dereddened for $E_{B-V}=0.8$. It shows the cooling of 1996 at almost 
  constant $V$ magnitude, the ``standstill'' of the color index in 1997, 
  and the early decline of 1998. Two measurements are available 
  for the deep decline in 1998 (arrows). Subsequently, the brightness 
  recovered in late 1998 -- early 1999, before fading to magnitudes 
  below $V=20$. Since no $B$ observations are available for late 1999 and
  2000, the subsequent evolution is unknown. The 
  declines of 1998 and 1999 are similar to the red declines found in 
  R CrB stars. }
\end{center}
\end{figure}

It is generally accepted that the dimming of V4334 Sgr is caused by dust.
The most convincing evidence is that the flux integrated over all optical and
infrared wavelengths has remained fairly constant (Fig. 5; see also Duerbeck
et al. 2000). Another evidence results from the
analyses of spectra taken in 1997 April and 1998 August (Pavlenko, Yakovina 
and Duerbeck 2000; Pavlenko and Duerbeck 2000). While the first observation
took place at maximum light, the second observation was
made when the brightness had declined by $\Delta V\sim 1.8^{\rm m}$. 
The modelling of the spectral
features yields a surface temperature for August 1998 
that is only $\sim 250$ K below the temperature $T_{\rm eff} = 5500$~K
determined for the 1997 April spectrum. The observed energy distribution,
however, indicates that the 
reddening had increased by $\Delta E(B-V)=0.6$ above the interstellar value,
which here was assumed to be 0.7 and was found to be consistent with the 
spectral energy distribution of the previous year. Thus, (1) there is indeed 
circumstellar extinction from 1998 onward, and (2) the pseudo-photospheric 
temperature of V4334 Sgr did not undergo major changes between 1997 and 1998.
\begin{figure}
\begin{center}
\includegraphics[width=90mm,angle=270]{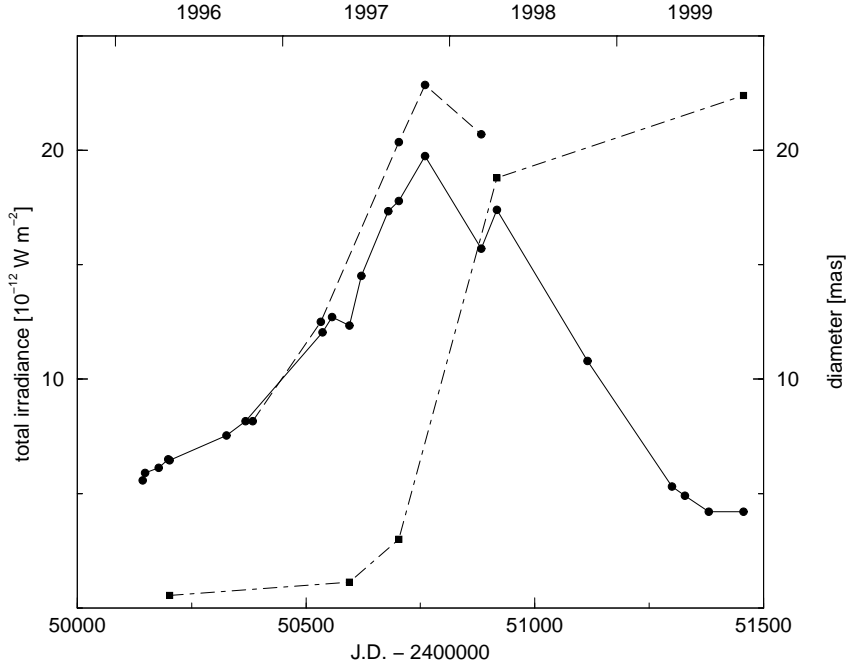}
\caption[]{Temporal evolution of the luminosity of V4334 Sgr and the diameter
  of its radiating dust shell. The circles connected by solid lines give the
  total irradiance shortward of $\rm 4.5~\mu m$, the circles connected with
  long dashes give the total irradiance up to $\rm 14.9~\mu m$ (ISO data). The
  squares connected by the dot-dashed line show the diameter of the shell in
  milliarcseconds (after Duerbeck et al. 2000).}
\end{center}
\end{figure}

After the massive dust formation phase of 1998 -- 1999, optical observations 
reveal only the ``tip of the iceberg''. While a pronounced IR excess 
had already evolved in 1997 -- 1998, the overwhelming percentage of the
radiation in 1999 -- 2000 is emitted at wavelengths in the mid-IR. Fig.~5
shows the integrated irradiance integrated over the optical and IR, as well as
the angular diameter of the IR emitting region as a function of time. 
Between early 1996 and early 1998, the total luminosity increased by a factor
4. The following decline is caused by the fact that most of the radiation is
radiated longward of $\rm 4.5~\mu m$, and was not taken into account because
of lack of data. The rapid growth of the angular diameter at the beginning of
1998 is especially remarkable; it shows that at that time dust  
was forming effectively and ubiquitously in the extended shell.

Irradiances and angular diameters will finally be converted into luminosities
and linear diameters. Since the problem of the distance of V4334 Sgr is not yet
settled (see the other contributions to this workshop),
I will refrain from deriving numbers that may be obsolete soon.

\section{Comparison with other final-helium-flash objects and lessons for the
  future}

Until a few years ago, only one final helium flash
object, FG Sge, was both well-studied and fairly well understood. 
Its evolution is very slow. V4334 Sgr evolves at a much faster pace, 
and has already overtaken FG Sge in its evolutionary stage (unless we are
observing different parts of the final helium flash evolution, see 
Lawlor and MacDonald 2000). Its photometric and spectroscopic 
evolution is similar to that of V605 Aql, the final helium flash object 
of 1919 (Duerbeck et al. 2000, see also Duerbeck et al. 2001). V4334 Sgr,
however, appears to evolve 50\% faster. It is important to follow 
the evolution of FG Sge, V4334 Sgr and V605 Aql, in order to learn more 
about the characteristics of final helium flash objects and their 
connection with other late stages of stellar evolution, like the 
R CrB stars, the PG1159 stars, and the non-DA white dwarfs. 

\section*{Acknowledgments}
This work would never have been possible without the help of various
colleagues. First, my wife, Waltraut Seitter let me participate in her 
researches on V605 Aql, the ``older twin'' of V4334 Sgr. Second, Arnout van
Genderen enthusiastically accepted my suggestion to include this object into
the Dutch telescope monitoring program, and many students from Leiden and
Groningen followed its photometric evolution over the years. Even more
instrumental for the success of this project was William Liller, who 
followed the evolution of Sakurai's object faithfully with his 0.2\,m 
telescope in Re$\rm \tilde{n}$aca, Chile. Martin Asplund, Stefano Benetti, 
Alfred Gautschy, Yakiv Pavlenko, Chris Sterken, and Patricia A. Whitelock 
added theoretical and 
observational expertise. Chris Sterken, Ronald Mennickent and Cath\'erine 
Delahodde carried out the observations in 2000. This research was supported 
by the Flemish Ministry for Foreign Policy, European Affairs, Science and 
Technology.

\end{article}

\begin{thebibliography}{}
\bibitem[]{}
Arkhipova, V.P., Noskova, R.I., Esipov, V.F., Sokol, G.V. 1999, Astr. Letters,
25, 615
\bibitem[]{}
Bath, G.T., Harkness, R.P. 1989, in Classical Novae, ed. M.F. Bode and
A. Evans (Chichester and New York: Wiley), p. 61
\bibitem[]{}
Duerbeck, H.W., Benetti, S., ApJ, 468, L111
\bibitem[]{}
Duerbeck, H.W., Benetti, S., Gautschy, A., van Genderen, A.M., Kemper, C.,
Liller, T., Thomas, T. 1997, AJ, 114, 1657
\bibitem[]{}
Duerbeck, H.W., Liller, W., Sterken, C., Benetti, S., van Genderen, A.M.,
Arts, J., Kurk, J.D., Janson, M., Voskes, T., Brogt, E., Arentoft, T., van der
Meer, A., Dijkstra, R. 2000, AJ, 119, 2360
\bibitem[]{}
Duerbeck, H.W., Hazen., M.L., Misch, A.A., Seitter, W.C. 2001 (this volume)
\bibitem[]{}
Guinan, E.F., Dewarf, L.E., McCook, G.P., Ditura, P., Mittal, R., Margheim,
S.J. 1998, BAAS, 193, 15.10
\bibitem[]{}
Iben, I., Jr., Kaler, J.B., Truran, J.W., Renzini, R. 1983, ApJ, 264, 605
\bibitem[]{}
Iben, I., Jr., MacDonald, J. 1995, in White Dwarfs, eds. D. Koester \&
K. Werner, (Berlin: Springer), 48
\bibitem[]{}
Iben jr., I., Tutukov, A.V., Yungelson, L.R. 1996, ApJ, 456, 750
\bibitem[]{}
Kerber, F., Palsa, R., K\"oppen, J., Bl\"ocker, T., Rosa, M.R. 2000, ESO
Messenger, No. 101, 27
\bibitem[]{}
Kimeswenger, S. 2001 (this volume)
\bibitem[]{}
Lawlor, T.M., MacDonald, J. 2000, preprint
\bibitem[]{}
Nakamo, S., Sakurai, Y., Hazen, M., Benetti, S., Duerbeck, H.W., 1996, 
IAU Circ. 6322
\bibitem[]{}
Pavlenko, Ya., Duerbeck, H.W. 2000, A\&A (in press)
\bibitem[]{}
Pavlenko, Ya.V., Yakovina, L.A., Duerbeck, H.W. 2000, A\&A, 354, 229 
\bibitem[]{}
Takamizawa, K. 1997, VSOLJ Variable Star Bull. 25, 4
\end{thebibliography}
\end{document}